# DFT insights into new B-containing 212 MAX phases: Hf$_2$AB$_2$ (A = In, Sn)

M. A. Ali[a,1], M. M. Hossain[a], M. M. Uddin[a], A. K. M. A. Islam[b,c], D. Jana[d], S. H. Naqib[b,2]

[a]Department of Physics, Chittagong University of Engineering and Technology (CUET), Chattogram-4349, Bangladesh
[b]Department of Physics, University of Rajshahi, Rajshahi-6205, Bangladesh
[c]Department of Electrical and Electronic Engineering, International Islamic University Chittagong, Kumira, Chittagong, 4318, Bangladesh
[d]Department of Physics, University of Calcutta, 92 A.P.C. Road, Kolkata, 700009, India

ABSTRACT

212 MAX phase borides are new additions to the MAX phase family with enhanced thermo-mechanical properties. In this article, first-principles calculations have been carried out to explore the mechanical properties, Vickers hardness, elastic anisotropy, thermal properties, and optical properties of predicted thermodynamically stable MAX compounds Hf$_2$AB$_2$ (A = In, Sn). The structural properties are compared with the available data to assess the validity of calculations. The mechanical stability of Hf$_2$AB$_2$ (A = In, Sn) compounds is established with the help of the computed stiffness constants ($C_{ij}$). The possible reason for enhanced mechanical properties and Vickers hardness of Hf$_2$SnB$_2$ is explained based on the analysis of bonding strength, followed by the electronic density of states. Higher mechanical strength and Vickers hardness of Hf$_2$AB$_2$ (A = In, Sn) compared to Hf$_2$AC (A = In, Sn) are also indicated in the light of charge density mapping. The values of Pugh ratio, Poisson's ratio and Cauchy pressure predict brittle character of the studied compounds. Besides, the anisotropic nature of the titled borides is investigated by 2D and 3D plots of elastic moduli along with some well established anisotropy indices. Thermal properties were investigated by calculating the Debye temperature, minimum thermal conductivity, Grüneisen parameter, and melting temperature. The thermal properties of Hf$_2$AB$_2$ (A = In, Sn) are also superior to Hf$_2$AC (A = In, Sn). The optical constants such as real and imaginary parts of the dielectric function, refractive index, extinction coefficient, absorption coefficient, photoconductivity, reflectivity, and loss function are investigated.

## 1. Introduction

The physical properties of MAX phases are always inspiring materials scientists to look for new MAX phases with better tunable properties for practical applications. The basis that makes the MAX phases attractive is the distinctive combination of the properties of metals and ceramics [1–3]. Consequently, researchers have disclosed the applications MAX phases in various sectors. For example, they can be used as alternative to graphite at high temperatures, as heating

Corresponding authors: [1]ashrafphy31@cuet.ac.bd; [2]salehnaqib@yahoo.com

elements, high temperature foil bearings and other tribological components, gas burner nozzles, and as tool for dry drilling of concrete [2]. The term MAX usually represents layered ternary carbides and nitrides where M is an atom from the transition metals group, A is a representative from the group of IIIA or IVA and X represents either carbon or nitrogen. The layering scheme is determined by an integer n which defines the sub-group of MAX phases with the values of n = 1, 2, or 3 [1,2,4,5]. Though most of the known 211 phases and two 312 phases ($Ti_3SiC_2$ and $Ti_3GeC_2$) were reported by Nowotny *et al.* [6–8] the real growth of the MAX phases was triggered by Barsoum's group in the 1990s [1,9]. However, the MAX phases were confined within C/N as X elements for a long time either as compounds [10–19] or alloys/solid solutions [20–29]. Some reports on the unusual MAX phases such as $Mo_2Ga_2C$ [12,30], 321 MAX phases [31], and atomically layered and ordered rare-earth i-MAX phases [32,33] are also available.

Recently, attention has been drawn by boron-containing MAX phases (B as X elements) after the deposition of atomic layers of boron on various metal surfaces successfully [34,35]. Moreover, prospective applications of 2D boron and their borides in nanoelectronic devices have also revitalized the interest on B containing compounds [36–38]. Attempts were made from different perspectives. Some reports are available just on the replacement C/N by B [39–43]. A group of ternary layered borides analogous to MAX was designed by Ade *et al.*[44] where X element was replaced by B; these borides are crystallized in a different system (orthorhombic space groups) unlike MAX phases [45,46]. Their attempts resulted in an important class of layered compounds that have also drawn significant attention. Wang *et al.*[47] was inspired by the efforts of Ade *et al.*[44] to search B-containing MAX phases and found $Ti_2InB_2$ which crystallized in the hexagonal system with space group $P\bar{6}m2$ (No. 187). The compound was synthesized by solid-state reaction method. Consequently Miao *et al.* [48] performed a high-throughput structure search and predicted a series of Zr and Hf-based thermodynamically stable borides with the typical features of MAX phases. They also claimed that the M-B (M = Zr and Hf) bonding is stronger than that of Ti-B owing to the lower electronegativity of Zr and Hf (1.32 and 1.16, respectively) than that of Ti (1.38). The lower electronegativity however, leads to charge transfer from Zr/Hf to B atoms, enhancing the stability of $BM_6$ octahedron within these compounds.

Among the predicted boride containing compounds, we have selected $Hf_2AB_2$ (A = In, Sn) (212 MAX phase) for our current project. Miao *et al.*[48] have only predicted the thermodynamic stability with limited information of electronic properties such as the electronic density of states and electron localization functions which are insufficient for forecasting possible practical applications. Therefore, more information such as on mechanical properties, elastic anisotropy, Vickers hardness, charge density mapping, Fermi surface, thermal and optical properties are needed for predicting possible relevance in many technological applications. The study of mechanical properties, elastic anisotropy, and Vickers hardness provide us with some useful information regarding their use as structural components or in other relevant sectors. Some prior knowledge of thermal properties helps to select the materials for high-temperature applications. The possible use of materials in optoelectronic devices as well as in coating technology can be predicted from the study of energy dependent optical constants. Therefore, the study of the mentioned properties is of great significance and we have performed a systematic study of these for the $Hf_2AB_2$ (A = In, Sn) compounds in this paper. Furthermore, the obtained properties are also compared with those of $Ti_2InB_2$ as well as of $Hf_2AC$ (A = In, Sn) (211 MAX phase) compounds.

## 2. Theoretical method and computational details

The optimization of the unit cell and calculations of the mentioned physical properties were carried out in the framework of density functional theory as implemented in the CASTEP code [49,50]. The generalized gradient approximation (GGA) of the Perdew-Wang 91 version (PW91) [51] was used for exchange and correlation functional. The pseudo atomic calculations were performed for B $2s^2\ 2p^1$, In $4d^{10}\ 5s^2\ 5p^1$, Sn $5s^2\ 5p^2$, Hf $5d^2\ 6s^2$ electronic states. The energy cutoff of the plane waves is taken as 500 eV, and a $10 \times 10 \times 4$ Monkhorst-Pack uniform mesh is used for the k-point [52] sampling. Density mixing was used for the calculations. The optimization of atomic configuration was done by Broyden Fletcher Goldfarb - Shanno (BFGS) geometry optimization technique [53]. The tolerance levels were set as follows: self-consistent convergence tolerance = $5 \times 10^{-6}$ eV/atom, maximum force on the atom = 0.01 eV/Å, maximum displacement of atoms = $5 \times 10^{-4}$ Å, and a maximum stress = 0.02 GPa. The phonon dispersion curve (PDC) and related parameters have been calculated using the Density Functional Perturbation Theory (DFPT) based linear-response method [54]. The important optical constants of the $Hf_2AB_2$ (A = In, Sn) compounds have been computed by the use of complex dielectric function $\varepsilon(\omega) = \varepsilon_1(\omega) + i\varepsilon_2(\omega)$. The imaginary part $\varepsilon_2(\omega)$ of the dielectric function $\varepsilon(\omega)$ is calculated using the equation,

$$\varepsilon_2(\omega) = \frac{2e^2\pi}{\Omega\varepsilon_0} \sum_{k,v,c} |\psi_k^c|u.r|\psi_k^v|^2 \delta\left(E_k^c - E_k^v - E\right),$$

where $\Omega$ is the unit cell volume, $\varepsilon_0$ is the dielectric constant of the free space, $u$ is the vector defining polarization of the incident electric field and $r$ is the position vector, $\omega$ is the light frequency, $e$ is the electronic charge and $\psi_k^c$ and $\psi_k^v$ are the conduction and valence band wave functions at $k$, respectively. The sums over $k$, $v$ and $c$ are used for the sampling of Brillouin zone in the $k$ space, contribution from occupied valence band (VB) and unoccupied conduction band (CB), respectively. The real part $\varepsilon_1(\omega)$ of the dielectric function $\varepsilon(\omega)$ has been derived from the imaginary part $\varepsilon_2(\omega)$ by using the Kramers–Kronig relation. Other optical constants: refractive index, extinction coefficient, absorption spectrum, reflectivity and energy-loss spectrum are calculated from the real and imaginary parts of the dielectric function using equations that can be found elsewhere [55].

## 3. Results and discussion

### *3.1 Structural properties and dynamical stability*

The unit cell structure of $Hf_2InB_2$ (212 MAX phase) is presented in Fig.1 which crystallizes in the hexagonal structure with the space group of $P\bar{6}m2$, (No. 187) [48]. The unit cell of $Hf_2InC$ is also given herewith to illustrate the difference between the structures. A 2D layer of B is squeezed in between two Hf layers [Fig. 1 (a)] that extensively contributes to the improved stability and strength of the structure. The covalent B-B bonding present in the boron layer is much stronger than the M-X bonding as present in the conventional C/N containing MAX phases, resulting in more stable structure of the B containing compounds of interest. For $Hf_2InB_2$, there are two types of B atoms, one is positioned at 1f Wyckoff site (0.6667, 0.3333, 0.5) and the other is positioned at 1b Wyckoff site (0.0, 0.0, 0.5). The Hf atom is located at 2h Wyckoff site (0.3333, 0.6667, 0.6935), and In/Sn atom occupies the 1e Wyckoff site (0.6667, 0.3333, 0.0). The obtained results for cell configuration are listed in Table 1. The results obtained by Miao et al.[48] are also presented in this table to weigh up the reliability of our calculated results. As seen from Table 1, our calculated values of $a$, $c$, and $V$ are consistent with the earlier reported values. An insignificant and/or tolerable level of deviations (0.64% (0.12%), 0.52% (0.25%) and 1.82% (1.07%) for *a*, *c*, and *V*, respectively of $Hf_2InB_2$ ($Hf_2SnB_2$) are observed, indicating the accurate settings of the parameters used for the calculations of the present study.

**Table 1:** Computed lattice parameters (*a* and *c*), *c/a* ratio, and volume (*V*) of $Hf_2AB_2$ (A = In, Sn) MAX phases, together with those of $Ti_2InB_2$.

| Phase | *a* (Å) | % of Variation | *c* (Å) | % of deviation | *c/a* | *V* (Å$^3$) | % of deviation | Reference |
|---|---|---|---|---|---|---|---|---|
| $Hf_2InB_2$ | 3.2206 | 0.64 | 8.4016 | 0.52 | 2.608 | 75.467 | 1.82 | This work |
|  | 3.200 |  | 8.358 |  | 2.611 | 74.119 |  | Theo. [48] |
| $Hf_2SnB_2$ | 3.2341 | 0.12 | 8.2248 | 0.25 | 2.543 | 74.503 | 1.07 | This work |
|  | 3.221 |  | 8.204 |  | 2.547 | 73.711 |  | Theo. [48] |
| $Ti_2InB_2$ | 3.080 | 0.09 | 7.940 | 0.29 | 2.578 | 65.277 | 0.55 | This work |
|  | 3.077 |  | 7.917 |  | 2.572 | 64.921 |  | Expt. [47] |

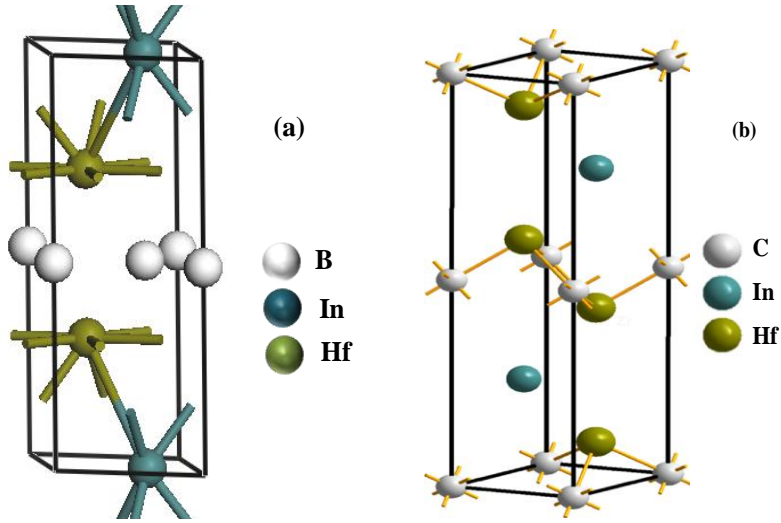

**Fig. 1:** The unit cell of (a) $Hf_2InB_2$ and (b) $Hf_2InC$.

After structural optimization, the dynamic stability of $Hf_2AB_2$ (A = In, Sn) MAX phases are confirmed by calculating phonon dispersion curves as shown in Fig. 2. As observed in Fig. 2, there is no imaginary phonon frequency in the whole Brillouin zone, indicating that $Hf_2AB_2$ (A = In, Sn) MAX phases are dynamically stable. Moreover, we have also checked the mechanical stability of these compounds presented in the following section. Noted that, Miao *et al.* [48] have shown their thermodynamical stability by calculating the phonon dispersion curve as well as formation energy.

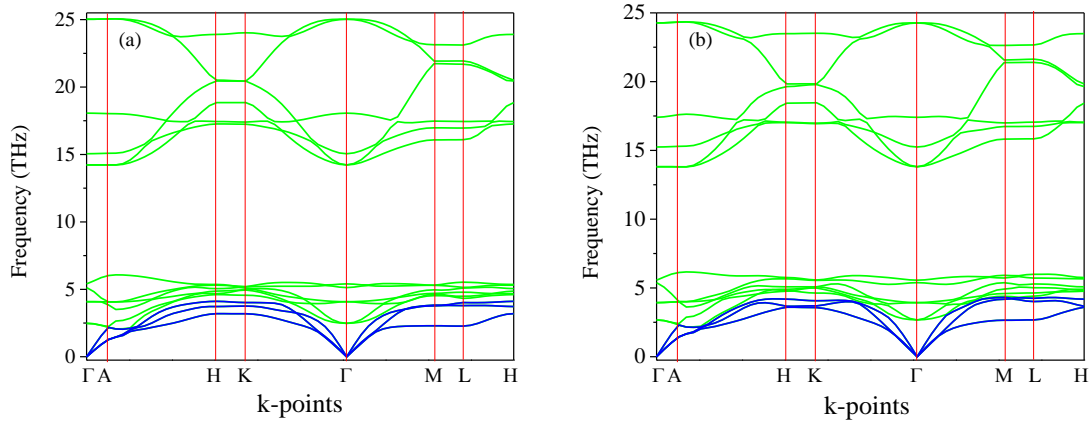

**Fig. 2:** Phonon dispersion curves of (a) $Hf_2InB_2$ and (b) $Hf_2SnB_2$.

## *3.2 Mechanical properties*

### *3.2.1 Stiffness constants and elastic moduli*

The study of stiffness constants ($C_{ij}$) of solids makes available a relationship between the mechanical and dynamical properties and supply important information regarding the scenario of the forces present within the solids. Besides, these constants are widely used to predict the mechanical stability of solids; a pre-requisite information required essentially before their use in practical applications. We have calculated the elastic stiffness constants and elastic moduli of $Hf_2AB_2$ (A = In, Sn) compounds using the strain-stress method; which has been widely used [56–59]. The calculated results, obtained for the first time, are presented in Table 1, together with the values of $Ti_2InB_2$ for comparison. Moreover, the results for $Hf_2AC$ (A = In, Sn) carbides are also presented for comparative understanding of the mechanical properties of $Hf_2AB_2$ (A = In, Sn) MAX phase borides.

The mechanical stability of hexagonal system can be judged by evaluating the following Born conditions [60]: $C_{11} > 0$, $C_{33} > 0$, $C_{44} > 0$, $C_{11}-C_{12} > 0$, $(C_{11} + C_{12})C_{33} - 2C_{13}^2 > 0$. As seen from the values listed in Table 1, we can predict that the $Hf_2AB_2$ (A = In, Sn) compounds are mechanically stable. In addition, some important information can be found from stiffness constants. For example, $C_{11}$ measures the stiffness along the *a*-axis under a pressure applied along [100] direction whereas $C_{33}$ measures the same along the *c*-axis under a pressure applied along [001] direction. As seen, $C_{11} > C_{33}$ for both Hf$_2$AB$_2$ (A = In, Sn) compounds, indicating a higher pressure is required for plastic deformation along the *a*-axis compared to the *c*-axis. The unequal values ($C_{11} \neq C_{33}$) revealed the anisotropic nature in the bonding strength. $C_{44}$ is an important stiffness constant found to be a better hardness predictor [61] than other elastic constants and moduli. The listed values of $C_{44}$ disclose that Hf$_2$SnB$_2$ is predicted to be harder than Hf$_2$InB$_2$ and Ti$_2$InB$_2$.

**Table 2:** The elastic constants, $C_{ij}$ (GPa), bulk modulus, *B* (GPa), shear modulus, *G* (GPa), Young's modulus, *Y* (GPa), machinability index, $B/C_{44}$, Pugh ratio, *G/B*, Poisson ratio, v and Cauchy Pressure, *CP* (GPa) of Hf$_2$AB$_2$ (A = In, Sn), together with those of Hf$_2$AC (A = In, Sn), Ti$_2$InB$_2$ and Ti$_2$InC.

| Phase | $C_{11}$ | $C_{12}$ | $C_{13}$ | $C_{33}$ | $C_{44}$ | B | G | Y | $B/C_{44}$ | G/B | v | Cauchy Pressure | Reference |
|---|---|---|---|---|---|---|---|---|---|---|---|---|---|
| Hf$_2$InB$_2$ | 343 | 61 | 76 | 278 | 94 | 154 | 114 | 274 | 1.64 | 0.74 | 0.20 | -33 | This work |
| Hf$_2$InC | 309 | 81 | 80 | 273 | 98 | 152 | 105 | 256 | 1.55 | 0.69 | 0.21 | -17 | Ref-[62] |
| Hf$_2$SnB$_2$ | 353 | 65 | 86 | 306 | 110 | 165 | 124 | 297 | 1.50 | 0.75 | 0.20 | -45 | This work |
| Hf$_2$SnC | 251 | 71 | 107 | 238 | 101 | 145 | 87 | 218 | 1.43 | 0.60 | 0.25 | -30 | Ref- [63] |
| Ti$_2$InB$_2$ | 364 | 47 | 58 | 275 | 94 | 147 | 122 | 287 | 1.56 | 0.83 | 0.17 | -47 | This work |
|  | 358 | 52 | 57 | 270 | 93 | 145 | 119 | 280 | 1.56 | 0.82 | 0.17 | -41 | Ref- [64] |
| Ti$_2$InC | 284 | 62 | 51 | 242 | 87 | 126 | 100 | 236 | 1.45 | 0.79 | 0.18 | -25 | Ref- [64] |

Furthermore, the stiffness constants were used to calculate the elastic moduli such as bulk modulus (*B*) and shear modulus (*G*) of the compounds of interest using the Voigt [65] and Reuss [66] models with the Hill's approximation [67,68] as follows: $B = (B_V + B_R)/2$, where $B_V = [2(C_{11} + C_{12}) + C_{33} + 4C_{13}]/9$ and $B_R = C^2/M$; $C^2 = C_{11} + C_{12}C_{33} - 2C_{13}^2$; $M = C_{11} + C_{12} + 2C_{33} - 4C_{13}$. $B_V$ represents the upper limit of *B* (Voigt bulk modulus) and $B_R$ represents the lower limit of *B* (Reuss bulk modulus). Like *B*, average values of Voigt ($G_V$) and Reuss ($G_R$) were used to calculate *G* using the following equations: $G = (G_V + G_R)/2$, where $G_V = [M + 12C_{44} + 12C_{66}]/30$ and $G_R = \left(\frac{5}{2}\right)[C^2 C_{44} C_{66}]/[3B_V C_{44} C_{66} + C^2(C_{44} + C_{66})]$; $C_{66} = (C_{11} - C_{12})/2$. The Young's modulus (*Y*) and Poisson's ratio (*v*) were also calculated using their relationships with *B* and *G*: $Y = 9BG/(3B + G)$ and $v = (3B - Y)/(6B)$ [69,70]. The volume deformation and shape deformation are the pure elastic deformation [71]. The bulk modulus disseminates knowledge of the elastic response of solids under uniform hydrostatic pressure whereas the knowledge of rigidity against the plastic deformation can be known from the shear modulus. As seen from Table 2, Hf$_2$SnB$_2$ exhibits more resistance to volume as well as to shape deformations compared to Hf$_2$InB$_2$ and Ti$_2$InB$_2$. Another important polycrystalline elastic modulus is Young's modulus, used to predict the stiff nature of solids; indicates that $Hf_2SnB_2$ is stiffer compared to $Hf_2InB_2$ and $Ti_2InB_2$ borides. For comparison, we have listed the reported values of $Hf_2InC$ and $Hf_2SnC$ MAX phases. From the table, it is clear that the compounds have larger values of $C_{11}$, $C_{33}$, $C_{44}$, $B$, $G$, and $Y$ for $Hf_2SnB_2$ compared to $Hf_2SnC$. The reported values of these parameters for $Hf_2InC$ are closer to the values of $Hf_2InB_2$, whereas its $C_{44}$ is slightly larger than that of $Hf_2InB_2$. The reported values of $Hf_2InC$ are calculated using the LDA (exchange-correlation term) within DFT that usually give inflated values of the elastic constants. One expects lower values if those are calculated using GGA to model the exchange-correlation functional [72]. The physical reason behind this variation of mechanical properties lies within the structures of 212 MAX phases ($Hf_2AB_2$; A = In, Sn) and 211 MAX phases ($Hf_2AC$; A = In, Sn) as shown in Fig. 1. Unlike in conventional MAX phases, there is a layer of B (2D layer) in $Hf_2AB_2$. The covalent B-B bonding present in the boron layer is much stronger than the M-X bonding as present in the conventional MAX phases, resulting in an enhanced bonding strength within the $Hf_2AB_2$ (A = In, Sn) compared to the 211 MAX phase carbides. Therefore, larger values of the mentioned parameters are expected for $Hf_2AB_2$ (A = In, Sn). This statement will be further elaborated from the charge density mapping to be presented later.

The machinability index (MI) [$B/C_{44}$][73] is usually used to predict the performance of solids in tribological sector. From Table 2, it is found that $Hf_2InB_2$ is more machinable than $Hf_2SnB_2$ and $Ti_2InB_2$. The MI is lowest for $Hf_2SnB_2$ owing to its high $C_{44}$, as it is well known that the harder solids are less machinable.

### 3.2.2 The brittleness of $Hf_2AB_2$ (A = In, Sn)

Failure mode (ductile/brittle) of a material can be characterized theoretically by three performance indicators such as the Pugh ratio [$G/B$], Poisson's ratio [$v$], and Cauchy pressure of compounds. The $G/B$ ratio is usually used to distinguish the solids into two broad classes: ductile or brittle which depends on the critical value $G/B$ = 0.571 [85]. The larger value than 0.571 defines the solids as brittle materials otherwise they are ductile. The ratio $v$ is also used to identify the failure mode of solid with a critical value 0.26 [74]. A value lower than 0.26 defines the solids as brittle materials otherwise it is ductile materials. Another tool that is used to classify the solids into brittle or ductile is the sign of the Cauchy pressure [$CP$] [75]. The negative value of $CP$ defines the solids as brittle materials otherwise they are ductile. Based on all the three parameters as listed in Table 2, we can say that the studied compounds fall into the class of brittle materials.

### 3.2.3 Theoretical values of Vickers hardness

The capability of materials to resist the plastic deformation when loaded is judged by their hardness. Several factors such as bonding strength, the structure of crystal, microstructure, and defects, etc. contribute to the hardness. There are several methods to determine the hardness of solids experimentally as well as theoretically and the results are different for different methods applied. A method developed by Gou *et al.* [76] has been well received by the scientific community for calculation of the Vickers hardness [10,46,77–79]. This method is applicable for a partial metallic system based on the Mulliken atomic population. The Vickers hardness ($H_v$) of $Hf_2AB_2$ (A = In, Sn) has been calculated by considering geometrical averages of all bonds present in the compounds considered here. The bond hardness can be expressed as $H_v^\mu = 740(P^\mu - P^{\mu'})(v_b^\mu)^{-5/3}$, where $P^\mu$ and $P^{\mu'}$ are Mulliken overlap population of the $\mu$-type bond and metallic population, respectively. The $P^{\mu'}$ is calculated as follows: $P^{\mu'} = \frac{n_{free}}{V}$; $n_{free} = \int_{E_P}^{E_F} N(E)dE$, where $E_P$ and $E_F$ are the energy at pseudogap and Fermi level, respectively. The $v_b^\mu$ is defined as volume of the $\mu$-type bond that is calculated as follows: $v_b^\mu = (d^\mu)^3/\sum_v[(d^\mu)^3 N_b^v]$. Finally, the Vickers hardness is calculated taking the geometric average of the bonds present in the compound using the equation: $H_V = \left[\prod \pi (H_v^\mu)^{n^\mu}\right]^{1/\sum n^\mu}$, where $n^\mu$ is the number of $\mu$-type bonds.

The obtained values of $H_v$ of $Hf_2AB_2$ (A = In, Sn) are listed in Table 3. Like the elastic moduli, the $H_v$ of $Ti_2InB_2$ (~ 4.05 GPa) as reported by Ali *et al.* [64] lies in between 3.94 GPa ($Hf_2InB_2$) and 4.41 GPa ($Hf_2SnB_2$). The Vickers hardness is also greater for $Hf_2AB_2$ (A = In, Sn) compared to those of $Hf_2AC$ (A = In, Sn). The reported values of $H_v$ for $Hf_2InC$ and $Hf_2SnC$ are 3.45 GPa [10] and 3.80 GPa [80], respectively which are lower than the values of $H_v$ for $Hf_2InB_2$ and $Hf_2SnB_2$. The same reason as explained in previous section for the difference in the values of elastic moduli is assumed to be responsible for this difference in the values of the Vickers hardness. The existence of much stronger B-B bonding can also be observed from the values of the individual bond hardness in Table 3. Ali *et al.* [64] also reported the higher hardness values for Ti$_2$InB$_2$ (4.05 GPa) compared to Ti$_2$InC (2.60 GPa).

**Table 3:** Computed Mulliken bond number $n^\mu$, bond length $d^\mu$, bond overlap population $P^\mu$, metallic population $P^{\mu'}$, bond volume $v_b^\mu$, bond hardness $H_v^\mu$ of $\mu$-type bond and Vickers hardness $H_v$ of Hf$_2$AB$_2$ (A = In, Sn).

| Compounds | Bond | $n^\mu$ | $d^\mu$ (Å) | $P^\mu$ | $P^{\mu'}$ | $v_b^\mu$ (Å$^3$) | $H_v^\mu$ (GPa) | $H_v$ (GPa) |
|---|---|---|---|---|---|---|---|---|
| Hf$_2$InB$_2$ | B1-B2 | 1 | 1.8594 | 2.38 | 0.0522 | 6.983591 | 67.51128 | 3.94 |
|  | B1-Hf1 | 2 | 2.5072 | 0.35 | 0.0522 | 17.12093 | 1.937556 |  |
|  | B2-Hf2 | 2 | 2.5072 | 0.32 | 0.0522 | 17.12093 | 1.742358 |  |
| Hf$_2$SnB$_2$ | B1-B2 | 1 | 1.8672 | 2.36 | 0.00199 | 6.928583 | 69.29496 | 4.41 |
|  | B1-Hf1 | 2 | 2.5131 | 0.35 | 0.00199 | 16.89358 | 2.315311 |  |
|  | B2-Hf2 | 2 | 2.5131 | 0.32 | 0.00199 | 16.89358 | 2.115716 |  |

Overall, the mechanical properties of the Hf$_2$AB$_2$ (A = In, Sn) are stronger than those of Hf$_2$AC (A = In, Sn), in good agreement with the findings of Wang *et al.* [47]. In 211 MAX phases, there is a C/N atom in the unit cell whereas there are two times B atoms in the unit cell of 212 MAX borides. This additional B atom positioned at the interstitial site forms B-B covalent bonding with surrounding B atoms within the B layer that enhanced the bonding strength significantly, and hence the enhanced mechanical properties and hardness for 212 MAX phase borides are observed. With these data in our hand, we now move on to discuss the electronic properties the 212 MAX borides.

## *3.3 Electronic energy density of states, charge density distribution and Fermi surface of Hf$_2$AB$_2$ (A = In, Sn)*

As seen in Tables 2 and 3, the mechanical properties and hardness of Hf$_2$SnB$_2$ are higher compared to those of Hf$_2$InB$_2$. Now the question is: why does this variation occur? To answer this question the electronic density of states (DOS) of Hf$_2$AB$_2$ (A = In, Sn) are calculated and shown in Fig. 3. The DOS is a useful tool to demonstrate the variation of mechanical properties and hardness among the series of solids where only one element remains as a variable, e.g., Sn replacing In when we move from Hf$_2$InB$_2$ to Hf$_2$SnB$_2$. As demonstrated by Fig. 3 that the DOS of Hf$_2$InB$_2$ and Hf$_2$SnB$_2$ are largely identical, except in the difference in the positions of the peaks. To illustrate the shift of the peaks, three red lines are used as references. For Hf$_2$SnB$_2$, the peaks are observed to be shifted to the lower energy regions with respect to those in Hf$_2$InB$_2$.

The bonding strength among the hybridized energy states can be realized from the positions of the peaks. The peaks shown in the DOS resulted in from the hybridization among the different states, e.g., hybridization between Hf-*d* states and In/Sn-*p* electronic states as shown in Fig. 3 (a) for $Hf_2InB_2$. The In/Sn-*p* states and B-*p* states also hybridize within these borides as can be found from the orbital resolved partial density of states [Fig. 3 (a) for $Hf_2InB_2$]. In general, the lower the hybridized energy states, the stronger is the bonding strength. Therefore, the peaks in the lower energy region promote stronger bondings, particularly the strong covalent bond formation. Therefore, the enhanced mechanical properties and Vickers hardness of $Hf_2SnB_2$ in comparison to $Hf_2InB_2$ as seen in Table 2 and Table 3 are consistent with the electronic DOS profiles.

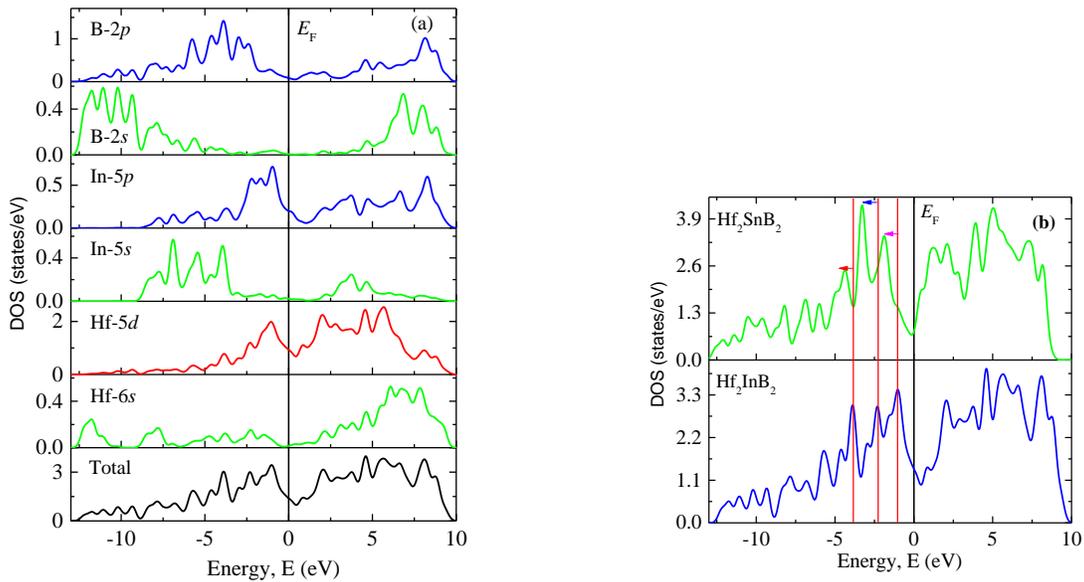

**Fig. 3:** (a) Total and partial DOS of $Hf_2InB_2$ and total DOS of (b) $Hf_2AB_2$ (A = In, Sn) measured with respect to Fermi energy (taken at 0 eV).

Moreover, it is seen from Table 2 that the mechanical properties of $Hf_2AB_2$ (A = In, Sn) are stronger than those of the corresponding 211 carbides. The Vickers hardnesses of titled borides are also higher than those of the corresponding 211 carbides. This variation is assumed due to the presence of strong covalent B-B bonding within these borides.

To illustrate the presence of the strong covalent B-B bonding, we have calculated the charge density mapping (CDM) as shown in Fig. 4. The CDM is a useful tool to demonstrate the presence of different bondings in terms of electronic charge density distribution in various crystal planes. The atomic positions are labeled to identify the accumulation and depletion of electronic charge. The adjoining scale specifies the high and low density of electronic charge with red and blue colors, respectively. The mapping of charge density includes and identifies areas with accumulation (high value) and depletion (low value) of electronic charges. As seen in Fig. 4, the charges are accumulated at the regions between B sites. Therefore, the strong covalent B-B bonding is expected by forming two center-two electrons (2c- 2e) bond for both compounds like $Ti_2InB_2$[47] and $Hf_3PB_4$[72]. For $Ti_2InB_2$, 0.87 |$e$|) charge is transferred from Ti atom to B atom, leading to the formation of 2c- 2e bonds between B atoms. For the present case, Mulliken analysis confirmed the transfer of charge 0.74 |$e$|) from Hf atoms to B atoms. The charge received from Hf atoms is shared among the B atom situated at the interstitials and the B atoms located at the edges; thus, 2c- 2e bond between B atoms within the 2D layer of B [Fig. 1 (a)] is formed. The individual bond hardness presented in Table 3 is also consistent with CDM results and our analysis based on elastic stiffness constants and moduli.

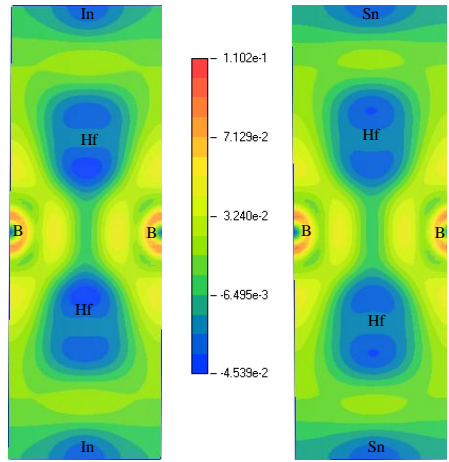

**Fig. 4:** Charge density mapping image of (left) $Hf_2InB_2$ and (right) $Hf_2SnB_2$.

Fig. 3 showing the DOS of $Hf_2AB_2$ (A = In, Sn) reveals the metallic nature of both compounds under study with finite values of DOS at the Fermi level. Band structure calculations (not shown) disclosed that three bands crossed the Fermi level for each of the two borides. Fig. 5 (a) and (b) show the Fermi surfaces (FS) of $Hf_2InB_2$ and $Hf_2SnB_2$, respectively. The two Fermi surfaces are almost identical; both electron and holes like sheets are present. The major contribution to the Fermi surface of $Hf_2AB_2$ (A = In, Sn) is due to the lowly dispersive Hf-5$d$ states. As seen from the FSs, there is one electron-like band and two hole-like bands. The electron-like band reveals the electronic motion in the *c*-direction while the hole-like bands reveal that charge carriers are confined within basal plane with greater velocity. Thus, it can be concluded that the $Hf_2AB_2$ (A = In, Sn) are characterized by charge transport anisotropy. The electron type charges contributed to the transport properties along the *c*-axis whereas the hole type charge carriers contributed to the transport properties confined within the basal planes.

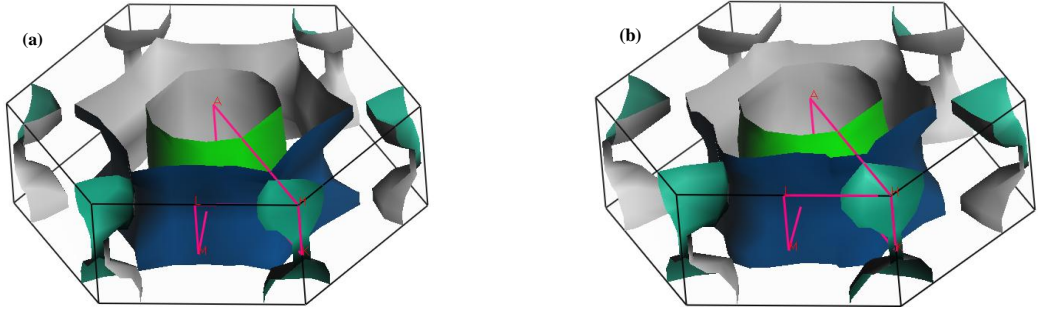

**Fig. 5:** Signature of anisotropic Fermi surface of (a) $Hf_2InB_2$ and (b) $Hf_2SnB_2$ compounds.

## *3.4 Mechanical anisotropy*

For practical applications of solids, like elastic properties themselves, knowledge about the anisotropic nature of these elastic properties are also vital. The mechanical anisotropy of solids is responsible for some important physical phenomena (crack behavior, anisotropic plastic deformation, and unusual phonon modes, etc.) [81][82]. The extent of anisotropy can be realized from the different values of elastic parameters in different crystallographic directions. Thus, the illustrations of Young's modulus, compressibility, shear modulus, and Poisson's ratio in the different crystal planes are computed using the ELATE code [83]; presented in Fig. 6 (a-d) for $Hf_2InB_2$ and Fig. 7 (a-d) for $Hf_2SnB_2$. The nature of anisotropy will be understood from the shape of the 3D plot. For isotropic solids, the 3D plots of these elastic parameters are expected to be perfectly spherical and their projections on different planes, circular. The deviation from the spherical and circular shapes represents the anisotropy. The anisotropic nature of both $Hf_2InB_2$ (Fig. 6) and $Hf_2SnB_2$ (Fig. 7) are similar except the values of the anisotropic indices. For example, both compounds have the maximum (minimum) values of *Y* along the horizontal (vertical) axis in *xz* and *yz* planes while *Y* is expected to be isotropic in *xy* plane as shown in Fig. 6 (a) and 7 (a). Fig. 6 (b) and 7 (b) show that the compressibility is maximum (minimum) at the vertical (horizontal) axis in *xz* and *yz* planes but seems to be isotropic in *xy* plane. A different anisotropic feature is observed for the shear modulus for both the compounds shown in Fig. 6 (c) and 7 (c). The minimum value of *G* is found along both axes in *xz* and *yz* planes with a maximum value at an angle of 45° in between the axes. Almost isotropic nature is observed in *xy* plane. The anisotropy of Poisson's ratio is illustrated in Fig. 6 (d) and 7 (d) where the maximum value is observed at the vertical axes in *xz* and *yz* planes with a minimum value at the horizontal axes for both the compounds. Like other parameters, the Poisson's ratio is isotropic in *xy* plane. The variations of the elastic moduli along the different directions are tabulated in Table 4.

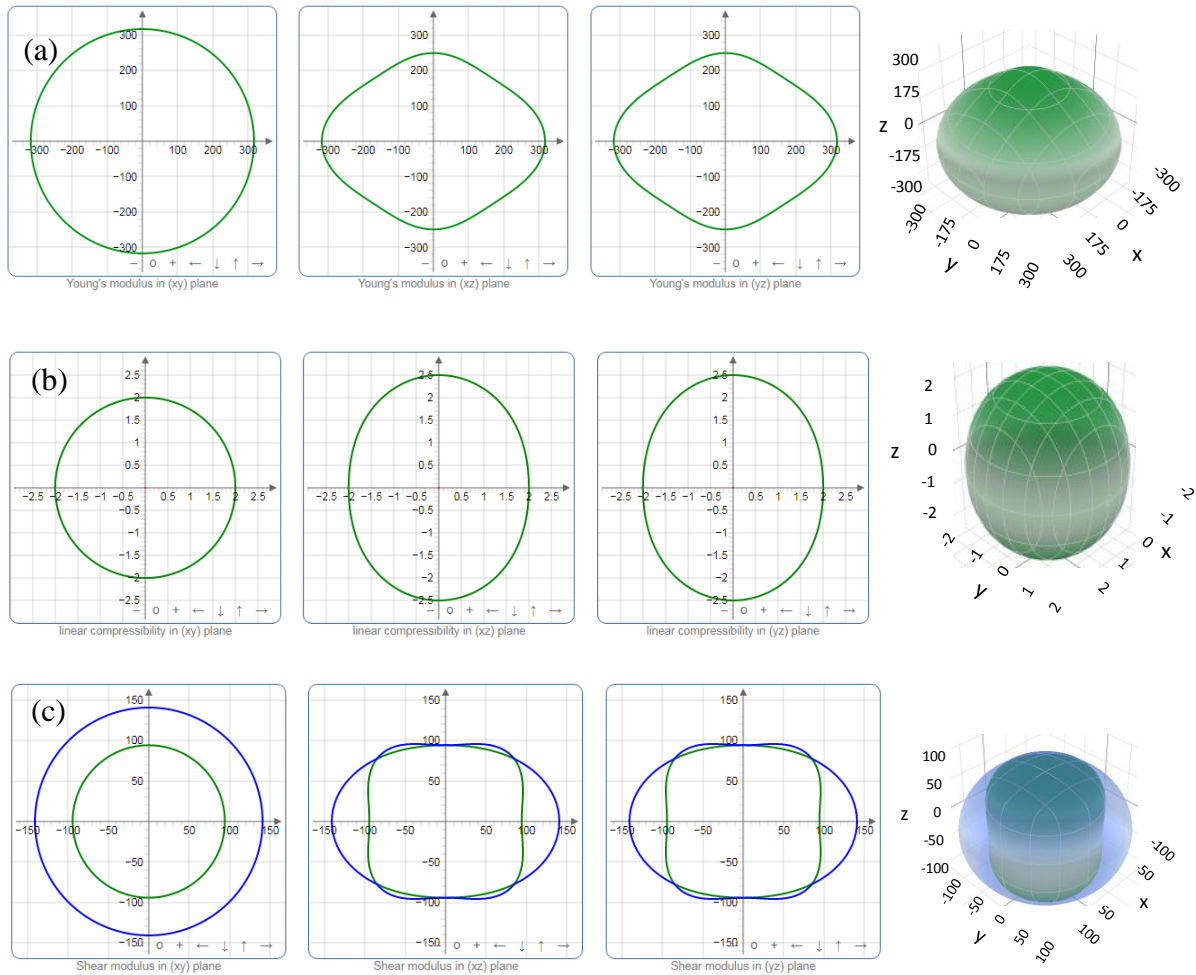

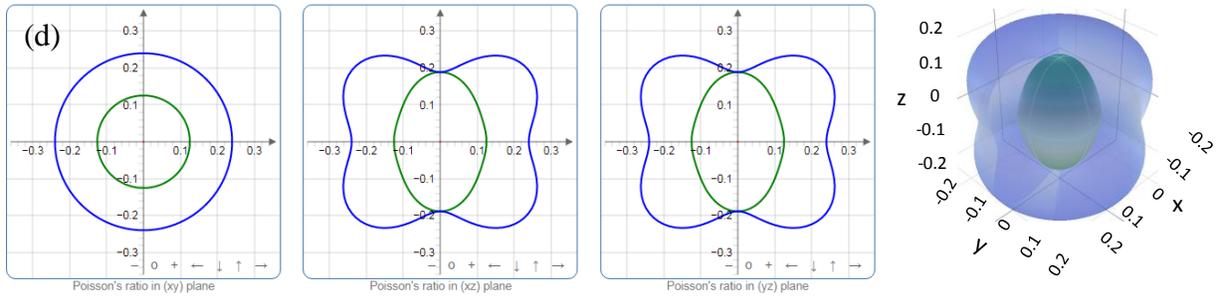

**Fig. 6:** The 2D and 3D plots of (a) $Y$, (b) $K$, (c) $G$ and (d) $v$ of $Hf_2InB_2$.

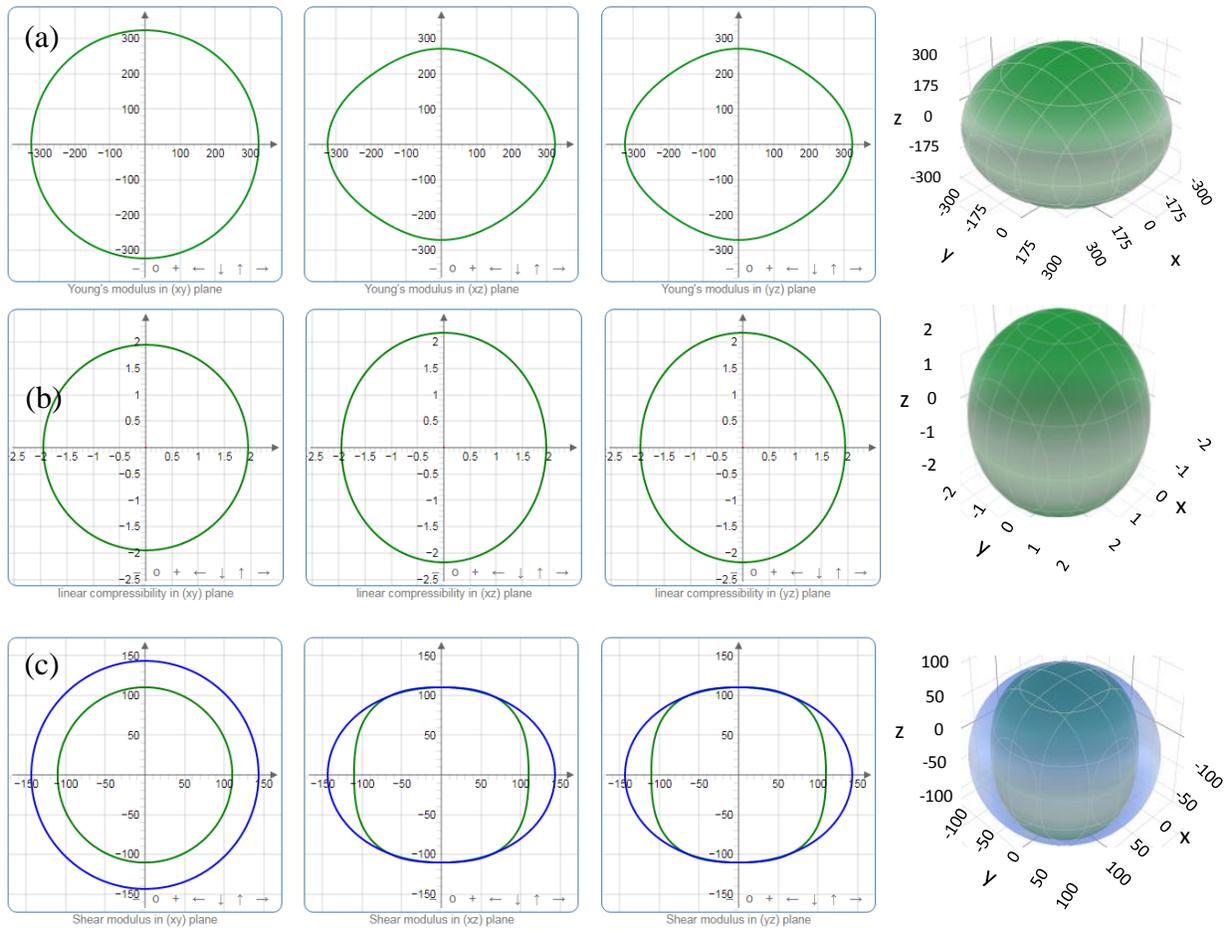

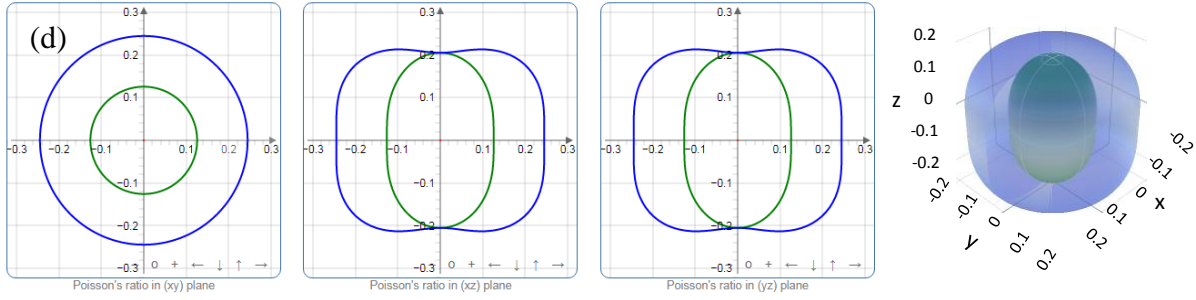

**Fig. 7:** The 2D and 3D plots of (a) $Y$, (b) $K$, (c) $G$ and (d) $v$ of $Hf_2SnB_2$.

**Table 4:** The minimum and the maximum values of the Young's modulus, compressibility, shear modulus, and Poisson's ratio of $Hf_2AB_2$ (A = In, Sn), together with those of the $Ti_2InB_2$ for comparison.

| Phase | $Y_{min.}$ (GPa) | $Y_{max.}$ (GPa) | $A_Y$ | $K$ ($TPa^{-1}$) | $K$ ($TPa^{-1}$) | $A_K$ | $G_{min.}$ (GPa) | $G_{max.}$ (GPa) | $A_G$ | $v_{min.}$ | $v_{max.}$ | $A_v$ |
|---|---|---|---|---|---|---|---|---|---|---|---|---|
| $Hf_2InB_2$ | 240.23 | 316.92 | 1.32 | 2.000 | 2.500 | 1.25 | 94.223 | 140.81 | 1.49 | 0.125 | 0.308 | 2.46 |
| $Hf_2SnB_2$ | 270.57 | 323.40 | 1.19 | 1.948 | 2.175 | 1.11 | 110.46 | 143.66 | 1.30 | 0.125 | 0.267 | 2.13 |
| $Ti_2InB_2$ | 240.80 | 249.11 | 1.45 | 2.042 | 2.775 | 1.36 | 094.13 | 158.89 | 1.69 | 0.095 | 0.309 | 3.25 |

Some other important anisotropy indices need to be explored owing to their scientific interest. The shear anisotropic factors $A_i$ (= 1, 2 and 3) are calculated using the relationships: $A_1 = \frac{\frac{1}{6}(C_{11}+C_{12}+2C_{33}-4C_{13})}{C_{44}}$, $A_2 = \frac{2C_{44}}{C_{11}-C_{12}}$, $A_3 = A_1 \cdot A_2 = \frac{\frac{1}{3}(C_{11}+C_{12}+2C_{33}-4C_{13})}{C_{11}-C_{12}}$ [84] for the {100}, {010} and {001} planes, respectively. The anisotropy index for bulk modulus along $a$ and $c$-directions is calculated using the following relations [85]: $B_a = a\frac{dP}{da} = \frac{\Lambda}{2+\alpha}$, $B_c = c\frac{dP}{dc} = \frac{B_a}{\alpha}$, where $\Lambda = 2(C_{11}+C_{12}) + 4C_{13}\alpha + C_{33}\alpha^2$ and $\alpha = \frac{(C_{11}+C_{12})-2C_{13}}{C_{33}+C_{13}}$. Finally, the anisotropy index for linear compressibility ($k_c/k_a$) along the $c$- and $a$-axis is calculated using the following relation [86]: $\frac{k_c}{k_a} = C_{11} + C_{12} - 2C_{13}/(C_{33} - C_{13})$. The obtained anisotropy factors are listed in Table 5. The obtained values of $A_i$'s ($A_i$ = 1 for isotropic nature); $B_a$ and $B_c$ ($B_a = B_c$ for isotropic nature) and $k_c/k_a$ ($k_c/k_a$ = 1 for isotropic nature) illustrated the anisotropic behavior of $Hf_2AB_2$ (A = In, Sn).

**Table 5:** Anisotropy factors, $A_1$, $A_2$, $A_3$, $k_c/k_a$, $B_a$, $B_c$ and universal anisotropy index $A^U$ of $Hf_2AB_2$ (A = In, Sn), together with those of $Ti_2InB_2$ for comparison.

| Phase | $A_1$ | $A_2$ | $A_3$ | $k_c/k_a$ | $B_a$ | $B_c$ | $A^U$ |
|---|---|---|---|---|---|---|---|
| $Hf_2InB_2$ | 1.16 | 0.67 | 0.78 | 1.25 | 499 | 400 | 0.177 |
| $Hf_2SnB_2$ | 1.04 | 0.76 | 0.79 | 1.12 | 514 | 460 | 0.040 |
| $Ti_2InB_2$ | 1.29 | 0.59 | 0.77 | 1.36 | 490 | 360 | 0.305 |

The universal anisotropy index $A^U$ [87] is calculated based on the upper limit (Voigt, $V$) and lower limit (Reuss, $R$) of bulk and shear modulus using the relation [87]: $A^U = 5\frac{G_V}{G_R} + \frac{B_V}{B_R} - 6 \geq 0$. The values of $A^U$ (non-zero) confirmed the anisotropic nature of the titled borides ($A^U = 0$, for isotropic solids). In summary, one can conclude that Hf$_2$AB$_2$ (A = In, Sn) compounds are anisotropic (Hf$_2$SnB$_2$ is less anisotropic) but their anisotropy indices are lower than those of Ti$_2$InB$_2$ (Table 4 and Table 5).

*3.4 Thermal properties*

The mechanical properties of solids can be correlated with the thermal properties via the Debye temperature ($\Theta_D$). Thus, it is of significant scientific interest to study the Debye temperature of Hf$_2$AB$_2$ (A = In, Sn) borides [29,88]. A well known method developed by Anderson [89,90] is used to calculate $\Theta_D$ using the equation: $\Theta_D = \frac{h}{k_B}\left[\left(\frac{3n}{4\pi}\right)N_A\,\rho/M\right]^{1/3} v_m$, where, $M$ be the molar mass, $n$ be the number of atoms in the molecules, $\rho$ be the mass density, and $h$, $k_B$, and $N_A$ be the Planck's constant, Boltzmann constant and Avogadro's number, respectively. $v_m$ is the average sound velocity which is calculated via the expression: $v_m = \left[1/3\left(1/v_l^3 + 2/v_t^3\right)\right]^{-1/3}$ where, $v_l$ and $v_t$ be the longitudinal and transverse sound velocities, respectively, that can be estimated from the shear modulus, bulk modulus, and density of solids by the following equations: $v_l = [(3B + 4G)/3\rho]^{1/2}$ and $v_t = [G/\rho]^{1/2}$. The obtained values of density, longitudinal, transverse and average sound velocities ($v_l$, $v_t$, and $v_m$, respectively) and Debye temperature ($\Theta_D$) are presented in Table 6. The value of $\Theta_D$ is higher for Hf$_2$SnB$_2$ (447 K) than that of Hf$_2$InB$_2$ (431 K) in accord with the mechanical properties. The values of $\Theta_D$ of Hf$_2$InC and Hf$_2$SnC are 383 K [62] and 393 K [91], lower than that of Hf$_2$InB$_2$ (431 K) and Hf$_2$SnB$_2$ (447 K); again consistent with the mechanical parameters and hardness values. The $\Theta_D$-value of Ti$_2$InB$_2$ is 633 K (slightly higher than the reported value of 621 K [64]), but much higher than those of Hf$_2$InB$_2$ and Hf$_2$SnB$_2$. The Hf-based MAX phases have low $\Theta_D$ values compared to other transition metal-based MAX phases. For instance, Bouhemadou *et al.* [92] have calculated the Debye temperature of Ti$_2$SC (800 K), Zr$_2$SC (603 K), and Hf$_2$SC (463 K) even though their elastic moduli are comparable. One of the possible reasons might be due to the higher atomic mass of Hf. The values of density, longitudinal, transverse, and average sound velocities of Hf$_2$AB$_2$ (A = In, Sn) compared to those of Ti$_2$InB$_2$ (presented in Table 6) are in favor of our explanation.

It is well known that the MAX phase materials have the prospect to be used at high-temperature applications. Therefore, the study of minimum thermal conductivity ($K_{min}$), Grüneisen parameter ($\gamma$), and melting temperature ($T_m$) of $Hf_2AB_2$ (A = In, Sn) is of immense importance owing to their possible relevance with the technology at high temperatures.

It is of scientific consequence to study the $K_{min}$ owing to the falling of the thermal conductivity of materials (e.g., ceramics) at high temperatures and eventually to a constant value ($K_{min}$) with the rise in temperature. $K_{min}$ is calculated using the equation [93] $K_{min} = k_B v_m \left(\frac{M}{n\rho N_A}\right)^{\frac{-2}{3}}$, where $k_B$, $v_m$, $N_A$ and $\rho$ are Boltzmann constant, average phonon velocity, Avogadro's number and density of crystal, respectively. The values of minimum thermal conductivity are given in Table 6 which are comparable with the $K_{min}$ of other MAX phases [24,39]. The value of $K_{min}$ for $Ti_2InB_2$ (1.19 (W/mK)) is slightly smaller than the reported value (1.23 (W/mK)) [64].

**Table 6:** DFT computed density ($\rho$), longitudinal, transverse and average sound velocities ($v_l$, $v_t$, and $v_m$, respectively), Debye temperature ($\Theta_D$), minimum thermal conductivity ($K_{min}$), Grüneisen parameter ($\gamma$) and melting temperature ($T_m$) of $Hf_2AB_2$ (A = In, Sn), together with those of $Hf_2AC$ (A = In, Sn), $Ti_2InB_2$ and $Ti_2InC$.

| Phase | $\rho$ (g/cm³) | $v_l$ (m/s) | $v_t$ (m/s) | $v_m$ (m/s) | $\Theta_D$ (K) | $K_{min}$ (W/mK) | $\gamma$ | $T_m$ (K) | Reference |
|---|---|---|---|---|---|---|---|---|---|
| $Hf_2InB_2$ | 10.86 | 5309 | 3240 | 3578 | 431 | 0.77 | 1.29 | 1800 | This work |
| $Hf_2InC$ | 11.67 | 5004 | 2999 | 3319 | 383 | | | 1691 | [62] |
| $Hf_2SnB_2$ | 11.08 | 5459 | 3344 | 3692 | 447 | 0.80 | 1.28 | 1872 | This work |
| $Hf_2SnC$ | 12.06 | 5121 | 3050 | 3376 | 393 | | | 1746 | [91] |
| $Ti_2InB_2$ | 05.90 | 7241 | 4545 | 5004 | 633 | 1.19 | 1.19 | 1858 | This work |
|  | 05.91 | 7168 | 4487 | 4942 | 621 | 1.23 | 1.20 | 1833 | [64] |
| $Ti_2InC$ | 06.08 | 6531 | 4055 | 4471 | 534 | 1.00 | 1.23 | 1569 | [64] |

The anharmonic effects within the $Hf_2AB_2$ (A = In, Sn) have been studied by calculating the Grüneisen parameter ($\gamma$) from the values of Poisson's ratio using the equation [94]: $\gamma = \frac{3}{2}\frac{(1+v)}{(2-3v)}$. The estimated values of $\gamma$ (Table 6) are within the expected range [0.85 to 3.53] for polycrystalline materials with the Poisson's ratio in the range of 0.05–0.46 [95]. Moreover, low anharmonic effects are expected owing to the low values of $\gamma$ within the titled borides. Another useful information of materials to predict their applications at high temperature is the melting temperature ($T_m$) that has been calculated for $Hf_2AB_2$ (A = In, Sn) using the equation [96], given by $T_m = 354 + \frac{4.5(2C_{11}+C_{33})}{3}$ and are listed in Table 6. The obtained values of $T_m$ (Table 6) are considerably large, in some cases higher than other Hf-based MAX phases (e.g., Hf$_3$SnC$_2$ ~1773 K) [97]. In fact, $T_m$ of Hf$_2$SnB$_2$ (1872 K) is higher than those of other 211 Hf-based MAX phases and only Hf$_2$PC (1845 K) and Hf$_2$PN (1861 K) have higher melting temperature than that of Hf$_2$InB$_2$ (1800 K).

From the PDC, the phonon DOS (not shown here) has been calculated for both compounds which is further used to calculate some thermodynamic properties such as Helmholtz free energy ($F$), energy ($E$), entropy ($S$), specific heat ($C_V$) and Debye temperature ($\Theta_D$) of Hf$_2$AB$_2$ (A = In, Sn) using the formalism described elsewhere [54]. Moreover, the linear thermal expansion coefficient (*TEC*), specific heat at constant pressure ($C_p$) are estimated according the following equations [98]:

$$\alpha = \frac{\gamma C_v}{3B_T v_m}$$

$$C_p = C_v(1 + \alpha \gamma T)$$

where, $B_T$, $v_m$ and $\gamma$ be the isothermal bulk modulus, molar volume and Grüneisen parameter, respectively.

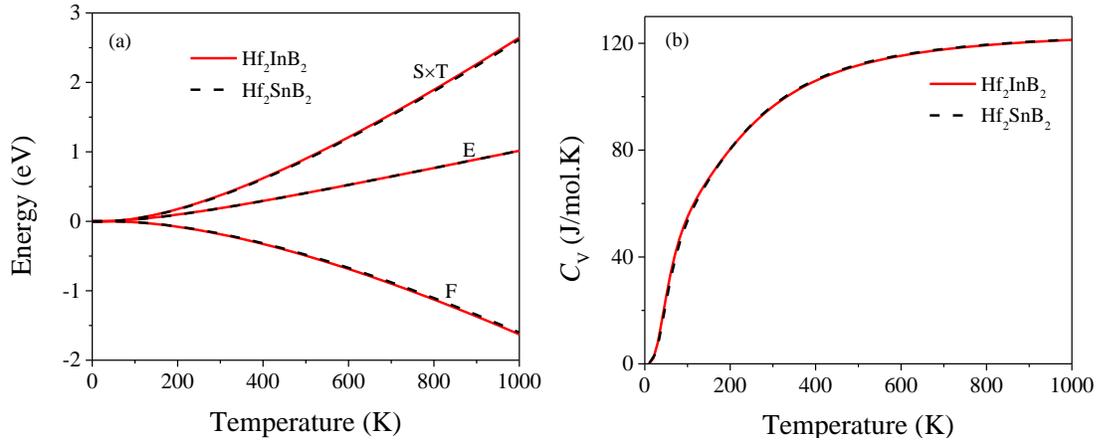

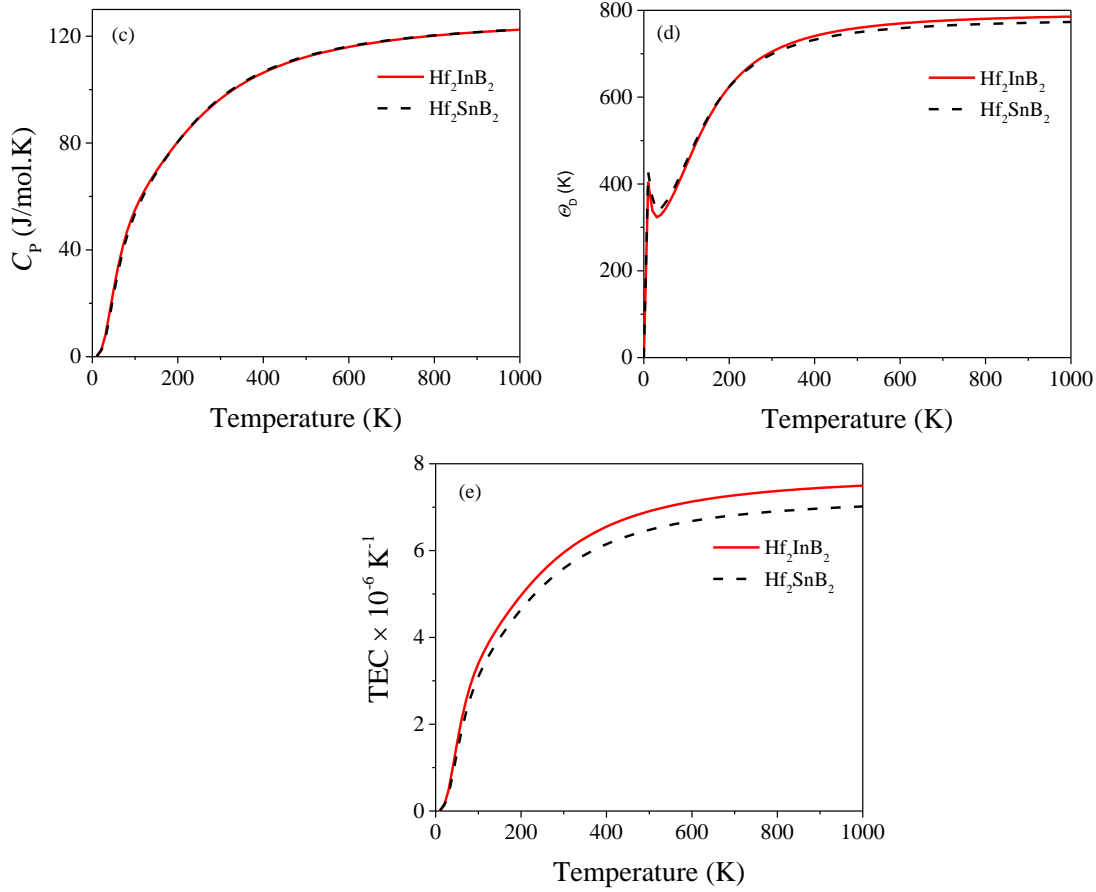

**Fig. 8:** The (a) Helmholtz free energy (*F*), energy (*E*), entropy (*S*), (b) specific heat ($C_V$), (c) specific heat ($C_P$), (d) Debye temperature, and thermal expansion coefficient (*TEC*) of $Hf_2AB_2$ (A = In, Sn) as a function of temperature.

The calculated *F*, *E* and *S* of $Hf_2AB_2$ (A = In, Sn) are shown in Fig. 8 (a) in the temperature range from 0 to 1000 K in which almost zero values of the parameters is observed at a temperature lower that 100 K. After that (above 100 K), the free energy is found to be decreased gradually with the increase of temperature, a common behavior of solids. While a reverse trend is observed for the entropy and enthalpy. The enhancement of disorder is expected owing to thermal disturbance; consequently, increase of entropy is observed in Fig. 8 (a). The enhancement of *E* with temperature rising is also expected for solids as shown in Fig. 8 (a).

Fig. 8 (b) and 8 (c) shows the temperature dependences of specific heats, $C_v$ and $C_p$ in the temperature range 0 -1000 K. The Debye $T^3$-law [99] holds good at low temperature. It is also indicative of the linear dispersion relation of phonon modes at long wave length limit. The increasing rate of specific heat is noted to be decreased with rise in temperature and expected to reach the classical Dulong–Petit (DP) limit ($C_V$ = 3nNk$_B$ = 125 J/mol-K) at high temperatures. Fig. 8 (d) also demonstrates the variation of Debye temperature with temperature as calculated from phonon DOS. As observed in Fig. 8 (d), increase of temperature leads to an increase of $\Theta_D$ which is a consequence of the changes in the vibrational frequency of atoms resulting from the increase in the thermal energy. The $\Theta_D$ and Vickers hardness of Hf$_2$InB$_2$ are higher than those of Hf$_2$InC, the $C_v$ of Hf$_2$InC reach the classical limit of DP law at lower temperature [10]; is in good agreement with the statement that lower $\Theta_D$ implies weaker solids and as a consequence, the heat capacity reach the classical DP value at relatively lower temperature [59]. Similar behavior was also reported for Ti$_2$InB$_2$ and Ti$_2$InC [64].

We have also calculated the thermal expansion coefficient (*TEC*) as function of temperature as shown in Fig. 8 (e). The *TEC* is observed to increase quickly particularly at temperature below 400 K, while the rate of increment becomes gradual at temperatures above 400 K; finally it attains a constant value at high temperature. The values of *TEC* for Hf$_2$InB$_2$ and Hf$_2$InB$_2$ are 5.90 × 10$^{-6}$ (K$^{-1}$) and 5.61 × 10$^{-6}$ (K$^{-1}$) at temperature ~ 300 K, respectively.

The low values of $K_{min}$, high melting temperature and moderate values of thermal expansion coefficient of Hf$_2$AB$_2$ (A = In, Sn) suggest that these compounds have potential to be used as thermal barrier coating (TBC) materials in high temperature technology.

*3.5 Optical properties*

To predict the possible use of Hf$_2$AB$_2$ (A = In, Sn) in optoelectronic devices and as absorbing or anti-reflection coating materials, the important optical constants are calculated. We have also calculated the optical constants of Ti$_2$InB$_2$ for comparison which are presented along with the results for the titled borides. The results imply that Hf$_2$AB$_2$ (A = In, Sn) compounds have potential to be used as alternative to Ti$_2$InB$_2$.

From Fig. 3, it is seen that the studied MAX compounds are metallic. Therefore, intra-band correction to the imaginary part of the dielectric constants is of importance. This is done by introducing the plasma frequency (3 eV) and a damping of 0.05 eV. Moreover, a smearing of 0.5 eV was used for Gaussian broadening for optical properties calculations. Total 13 numbers of bands are considered and the k-point mesh of 9 × 9 × 4 is selected for this calculation. The calculated optical properties are presented in Fig. 9 and Fig. 10. The energy range of incident photon energy is taken up to 25 eV and in general the [100] polarization direction of the electric field is chosen. Moreover, some values are calculated for both [100] and [001] directions to explore the anisotropic nature of the optical parameters.

The most general optical constant of materials is the dielectric function that describes the response of a material to the incident electromagnetic wave. Figs. 9 (a and b) show the real and imaginary part of the dielectric function. The peaks in $\varepsilon_2(\omega)$ are associated with the electron excitation. The Drude-like nature of the studied compounds is confirmed from the large negative value of $\varepsilon_1(\omega)$. A very sharp peak in $\varepsilon_2(\omega)$ for all the compounds is observed at very low energy region (~0.05 eV) owing to the intra-band transition of electrons within the Hf-$5d$ states.

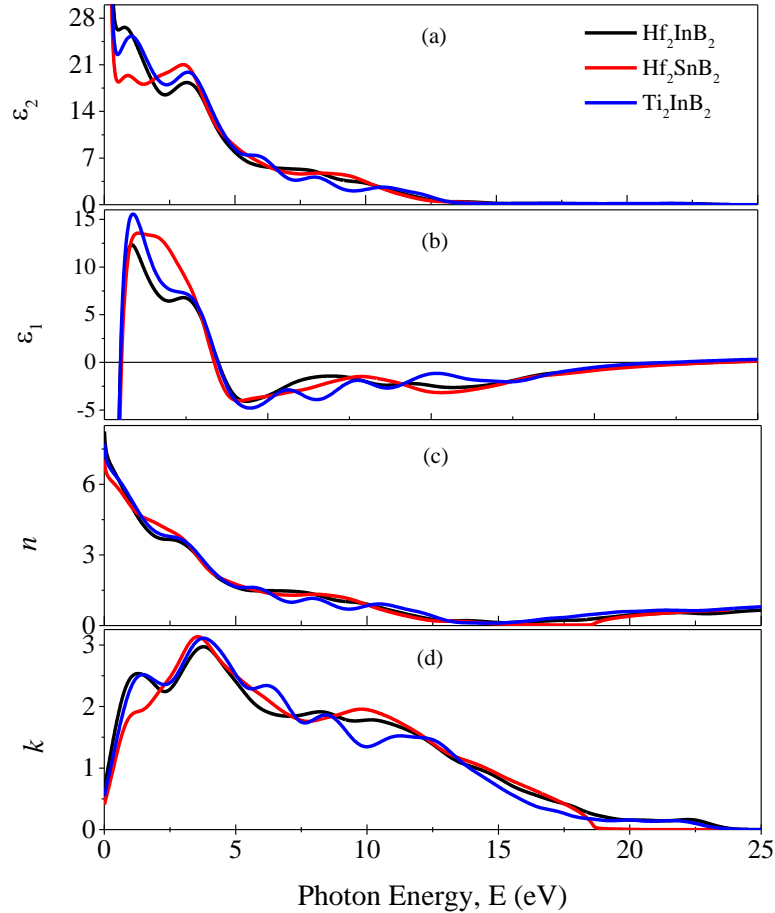

**Fig. 9:** The (a) real ($\varepsilon_1$), and (b) imaginary ($\varepsilon_2$) part of dielectric function, (c) refractive index ($n$) and (d) extinction coefficient ($k$) of Hf$_2$AB$_2$ (A = In, Sn), together with those of the Ti$_2$InB$_2$ for comparison.

The refractive index is a characteristics property of optical materials that has real part, $n$ (measures the phase velocity of incident radiation) and imaginary part, $k$ (measures loss due to absorption during the propagation of electromagnetic radiation through the solids). The static values of $n(0)$ are 8.24 (6.84), 7.03 (5.60) and 7.71 (7.72) for $Hf_2InB_2$, $Hf_2SnB_2$ and $Ti_2InB_2$, respectively for the [100] ([001]) polarization. The energy region in which $k(\omega)$ is higher than $n(\omega)$ is the region where light cannot pass through the solids without significant attenuation [100]. It is seen that the curves of $n(\omega)$ and $k(\omega)$ [Fig. 9 (c) and (d)] follow the pattern of $\varepsilon_1(\omega)$ and $\varepsilon_2(\omega)$, respectively.

The absorption coefficient ($\alpha$) measures the ability of materials to absorb incident radiation. Fig. 10 (a) shows the absorption spectra of $Hf_2AB_2$ (A = In, Sn) wherein a non-zero value of $\alpha$ at zero photon energy is seen owing to the metallic nature of the studied compounds, which is consistent with band structure and DOS studies. The spectra are observed to be increased with the increase in incident energy and in the spectral range 10 - 12.5 eV exhibited the strongest absorption region; decreasing thereafter with further increase of photon energy. Sometimes, very large light absorption in the visible and ultraviolet region is very helpful to design some devices like medical sterilizer machines. Fig. 10 (b) shows the photoconductivity ($\sigma$) of $Hf_2AB_2$ (A = In, Sn) wherein the metallic nature is seen from the non-zero values of $\sigma$ at the starting of the incident photon energy.

One of the most important applications of MAX phase materials is as coating materials to diminish solar heating. This ability can be judged by studying the reflectivity of target materials. We have calculated the reflectivity spectrum of $Hf_2AB_2$ (A = In, Sn) as shown in Fig. 10 (c). It was predicted by Li *et al.* [101,102] that the MAX phase materials with reflectivity greater than 44% will be able to reduce solar heating. For the present case, it can be judged that $Hf_2AB_2$ (A = In, Sn) compounds fulfill this criterion. Therefore, $Hf_2AB_2$ (A = In, Sn) compounds could be used as coating materials to reduce solar heating. Moreover, the values of $R(0)$ for both [100] and [001] polarization directions of the electric field are 0.6161and 0.5564 for $Hf_2InB_2$; 0.5650 and 0.4863 for $Hf_2SnB_2$; 0.5966 and 0.5961 for $Ti_2InB_2$, respectively.

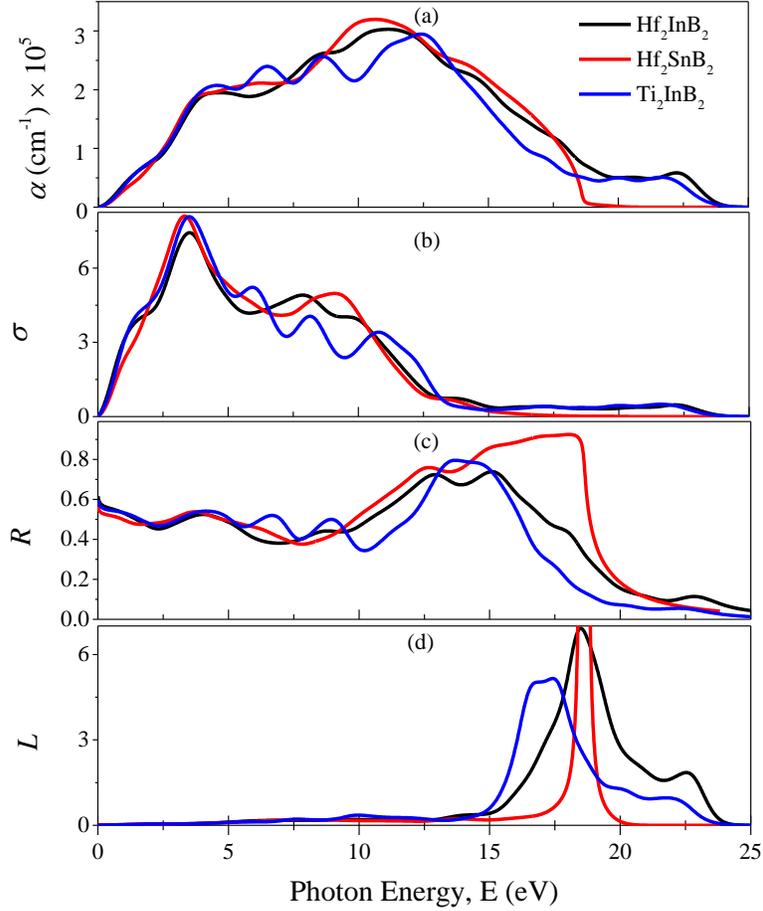

**Fig. 10:** The (a) absorption coefficient (α), (b) photoconductivity (σ), (g) reflectivity (R) and (h) loss function (L) of $Hf_2AB_2$ (A = In, Sn), together with those of the $Ti_2InB_2$ for comparison.

Finally, the loss function of $Hf_2AB_2$ (A = In, Sn) compounds are studied as shown in Fig. 10 (d). The spectra exhibit the highest peak at particular incident energy/frequency that is known as the plasma frequency ($\omega_p$) of solids. The $\omega_p$ for both [100] and [001] polarization directions are 18.5 eV and 18.81 eV for $Hf_2InB_2$; 18.62 eV and 18.86 eV for $Hf_2SnB_2$; 17.43 eV and 16.85 eV for $Ti_2InB_2$ respectively, where a swift decrease of reflectivity is observed. This is the characteristics frequency above which the materials behave as dielectrics. At the end of this section it should be mentioned that the values of $n(0)$, $R(0)$ and $\omega_p$ of $Hf_2AB_2$ (A = In, Sn) and $Ti_2InB_2$ are different for both [100] and [001] polarization directions, indicating the anisotropic nature of the optical parameters.

## 4. Conclusions

The recently predicted thermodynamically stable B-containing 212 MAX phase $Hf_2AB_2$ (A = In, Sn) ternary borides are investigated in details by the density functional theory. The calculated lattice constants are consistent with the available data. The mechanical stability of the $Hf_2AB_2$ (A = In, Sn) compounds is seen to be satisfied through Born conditions. The increase of elastic moduli and Vickers hardness is observed from $Hf_2InB_2$ to $Hf_2SnB_2$. Moreover, the mechanical properties and Vickers hardness of $Hf_2AB_2$ (A = In, Sn) borides are higher than those of $Hf_2AC$ (A = In, Sn) MAX phase carbides owing to the presence of strong covalent B-B bonding within the 2D boron layer. The charge density mapping confirms the strong covalent B-B bonding. Like $Hf_2AC$ (A = In, Sn), the brittleness as well as anisotropic nature of $Hf_2AB_2$ (A = In, Sn) is confirmed. The values of parameters concerning the thermal properties of $Hf_2SnB_2$ are higher than those of $Hf_2InB_2$, in good agreement with the results obtained for elastic and mechanical properties. Furthermore, the thermal parameters of these studied borides are also found to be higher than those of $Hf_2AC$ (A = In, Sn). This indicates better suitability of these 212 borides $Hf_2AB_2$ (A = In, Sn) for high-temperature applications. The values of $K_{min}$ (low values), $\Theta_D$ (high values) and *TEC* (moderate values) make sense for their possible use as thermal barrier coating materials. The titled borides have the potential to be used as coating materials to diminish solar heating. In comparison with $Ti_2InB_2$, $Hf_2SnB_2$ has enhanced mechanical properties, Vickers hardness, and melting temperature while $Hf_2InB_2$ has low values of the mentioned parameters compared to those of $Ti_2InB_2$. We hope that the results obtained herein for $Hf_2AB_2$ (A = In, Sn) MAX phase borides will encourage the materials scientists to try to synthesis these boron-containing 212 MAX phases and study their properties in greater depth both experimentally and theoretically in future.